\documentclass[12pt]{l4dc2021} 

\usepackage{bm}
\usepackage{mathtools}
\usepackage{dsfont}

\makeatletter
\let\Ginclude@graphics\@org@Ginclude@graphics
\makeatother

\usepackage{textcomp}

\newcommand*{\QED}{\hfill\ensuremath{\square}}%

\def\BibTeX{{\rm B\kern-.05em{\sc i\kern-.025em b}\kern-.08em
    T\kern-.1667em\lower.7ex\hbox{E}\kern-.125emX}}
    
\newcommand{\rknorm}[1]{\Vert #1 \Vert_\mathcal{H}}

\DeclareMathOperator*{\argmin}{\arg\,\min}  % rbp

\newcommand{\pushright}[1]{\ifmeasuring@#1\else\omit\hfill$\displaystyle#1$\fi\ignorespaces}

\definecolor{coolYellow}{HTML}{FDFD97}
\definecolor{coolGreen}{HTML}{77AC30}
\definecolor{coolBlue}{HTML}{0072BD}

\newtheorem{assumption}{Assumption}
\newtheorem{myRemark}{Remark}
\newtheorem{myProposition}{Proposition}
\newtheorem{myLemma}{Lemma}
\newtheorem{myTheorem}{Theorem}

\title[Deterministic error bounds for kernel-based learning techniques under bounded noise]{Deterministic error bounds for kernel-based \\ learning techniques under bounded noise}
\usepackage{times}
\author{%
 \Name{Emilio T. Maddalena$^1$} \Email{emilio.maddalena@epfl.ch}\\
 \Name{Paul Scharnhorst$^{1,2}$} \Email{paul.scharnhorst@epfl.ch}\\
 \Name{Colin N. Jones$^1$} \Email{colin.jones@epfl.ch}\\[3pt]
 \addr $^1$\'Ecole Polytechnique F\'ed\'erale de Lausanne, Lausanne, Switzerland\\
 \addr $^2$Swiss Center for Electronics and Microtechnology, Neuchâtel, Switzerland%
}

\begin{document}

\maketitle

\begin{abstract}%
We consider the problem of reconstructing a function from a finite set of noise-corrupted samples. Two kernel algorithms are analyzed, namely kernel ridge regression and $\varepsilon$-support vector regression. By assuming the ground-truth function belongs to the reproducing kernel Hilbert space of the chosen kernel, and the measurement noise affecting the dataset is bounded, we adopt an approximation theory viewpoint to establish \textit{deterministic}, finite-sample error bounds for the two models. Finally, we discuss their connection with Gaussian processes and two numerical examples are provided. In establishing our inequalities, we hope to help bring the fields of non-parametric kernel learning and system identification for robust control closer to each other.
\end{abstract}


\section{Introduction}

As opposed to classical system identification techniques, where the structure of a finite-dimensional model is chosen \textit{a priori}, kernel-based methodologies deal with possibly infinite-dimensional hypothesis spaces. In the latter, the number of parameters to be determined is not fixed, but depends on the number of available data-points. This non-parametric approach to building models is popular in many disciplines, usually in the form of Gaussian processes (GPs)|whose means are weighted sums of kernels|or plain radial basis functions (RBFs) \citep{jakhetiya2020kernel,moriconi2020high,singh2019kernel}. Among systems and control researchers, kernel methods have also been studied with the aim of adapting and improving existing tools (see \cite{pillonetto2014kernel,chiuso2019system,ljung2020shift} for recent reviews). For instance, an in-depth analysis of linear system identification through stable kernels, which encode the asymptotic decay of the plant impulse response, was carried out in \cite{pillonetto2010new,carli2016maximum,chen2018continuous}; see also the work \cite{lataire2016transfer} for a frequency domain perspective of the same problem, and \cite{blanken2020kernel} for the case of non-causal dynamical systems. Kernel-based identification of Hammerstein and Wiener systems, i.e. linear systems in cascade with a static nonlinearity, was studied from a similar point of view in \cite{risuleo2017nonparametric,risuleo2019bayesian}.

When nonlinear dynamics are considered in their full generality as in this work, a commonly adopted approach is to directly assume the availability of pairs of current and next states |that is, state-space models \citep{koller2018learning,bradford2020stochastic,umlauft2020learning}. The goal then is to reconstruct the complete vector field rather than the time-response of the unknown system (see \cite{pillonetto2011new} for an exception to this statement), a function reconstruction problem remarkably similar to the ones found in machine learning. Certainly, embedding prior knowledge regarding for instance stability can be quite challenging in this context \citep{umlauft2020learning}.

Exploiting machine learning techniques to model and subsequently control physical systems requires caution, especially in safety-critical applications. The fact that GPs provide users with not only nominal predictions, but also confidence intervals, is arguably one of the main reasons for their popularity \citep{bradford2019nonlinear,binder2019approximate,hewing2020learning}. This probabilistic uncertainty measure can then be used to assess the risk associated with actions, thus enabling the use of stochastic analysis techniques \citep{polymenakos2019safety,jackson2020safety}. If a non-probabilistic viewpoint is taken instead, it is possible to derive hard prediction-error bounds for the learned models such that the unknown ground-truth cannot lie outside the established `prediction envelope'. These types of guarantees are widely known in the field of approximation theory \citep{schaback2000unified,wendland2004,fasshauer2011positive,wang2019prediction}, where authors often consider the problem of interpolating noise-free observations. Whereas having access to perfect measurements might be common in domains such as computer graphics, it is definitely not the case in power networks, robotics, building automation systems, etc. Our goal herein is to extend such theory to the scenario where the designer is confronted with unknown but bounded noise, hence providing new tools for deterministic safety certification and robust control. Besides the latter application domain, the developed theory could also be of use in branches of natural sciences where the bounded noise assumption is regarded as more adequate than the Gaussian one \citep{d2013bounded}.

\textbf{Contributions:} In this note, we study the problem of learning an unknown real-valued function $f$ from a set of evaluations corrupted by noise. To achieve this goal, we employ two distinct non-parametric kernel techniques, namely the popular kernel ridge regression (KRR), and $\varepsilon$-support vector regression (SVR) \citep{schlkopf2018learning}. Two main assumptions are made\footnote{Note that these are fundamentally different compared to the ones made in the Gaussian processes setting. For more information, please refer to the discussion in Subsection~\ref{subsec:GPcomparison}.}: firstly, the measurement noise is bounded by a known finite quantity; secondly, given a kernel $k$, the unknown $f$ belongs to the reproducing kernel Hilbert space (RKHS) $\mathcal{H}$ of $k$, and an upper bound for its norm $\rknorm{f}$ is known. The same RKHS assumptions were also exploited in the works \cite{srinivas2012information,koller2018learning,umlauft2019feedback,hashimoto2020learning}. We then establish deterministic prediction-error bounds|otherwise known as risk or generalization error bounds|which extend classical results valid for noise-free interpolants only. Our final expressions are moreover given in closed-form, requiring only the solution of a simple box-constrained quadratic program in the KRR case. Finally, two numerical experiments are presented and a comparison with an existing alternative is provided.

\textbf{Notation:} $\mathbb{N} := \{1,2,\dots\}$, $\mathbb{R}^n$ is the $n$-dimensional Euclidean space, and $\mathbb{R}^n_{\geq 0}$ ($\mathbb{R}^n_{> 0}$) its positive (strictly positive) orthant. Given a matrix $A$, $A^\top$ denotes its transpose, and $A\succ 0$ indicates that $A$ is positive-definite. $I$ and $\mathbf{0}$ are respectively the identity and the zero matrices of appropriate sizes. Unless otherwise specified, $f$ denotes a map and $f(x)$, its particular evaluation at a point $x$. Given two vectors $a$ and $b$, the inequality $a \leq b$ and the absolute value $|a|$ are to be read element-wise. $\Vert a \Vert_1$ and $\Vert a \Vert_2$ will be used respectively for the 1-norm and 2-norm of a vector, whereas $\rknorm{f}$, for the RKHS functional norm.

% \begin{figure}[t]
%     \floatconts
%     {fig:example}% label
%     {\vspace{-20pt} \caption{Left: asdhasdhas.}}% 
%     {\includegraphics[scale=0.28]{pics/fig2a}
%     \hspace{1pt} \includegraphics[scale=0.28]{pics/fig2b}
%     \hspace{1pt} \includegraphics[scale=0.28]{pics/fig2c}}
%     \label{fig.PENDULUM}
% \end{figure}

\section{Problem setting}
\label{sec:prob}

Consider the problem of learning an unknown map $f:\mathcal{X} \to \mathbb{R}$, referred to as the ground-truth or target function. 
%\footnote{The more general case $f:\mathcal{X} \to \mathbb{R}^m$ can be handled by learning each output dimension separately.}. 
Herein $\mathcal{X}\subset \mathbb{R}^m$ is assumed to be compact. In order to reconstruct $f$, we collect a finite dataset
\begin{equation}
    D = \{(x_n,y_n) \, | \, n=1,\dots,N\}
\end{equation}
composed of sites $x_n$ and noisy evaluations of the ground-truth
\begin{equation}
\label{eq:genprob}
    y_n = f(x_n) + \delta_n, \ n=1,\dots, N
\end{equation}

\begin{assumption}
The data sites $X = \{ x_1,\dots,x_N \}$ in $D$ are pairwise distinct.
\label{as.pairwiseDistinct}
\end{assumption}

\begin{assumption}
The measurement noise $\delta = \begin{bmatrix}\delta_1 & \dots & \delta_N\end{bmatrix}^\top$ is bounded $|\delta| \leq \bar{\delta}$, with $\bar{\delta} \in \mathbb{R}^N_{\geq 0}$.
\label{as.noiseModel}
\end{assumption}

Working with bounded uncertainties is at the core of robust analysis and control. Similar assumptions were also made in recent learning-based techniques \citep{rosolia2017robust,novara2019nonlinear,manzano2020robust}. Next, we review basic definitions and results from RKHS theory, which are used as a starting point for Section~\ref{sec:crafting}. The reader is referred to \cite{manton2015primer} for a more in-depth discussion on the topic.

\subsection{Kernels and their RKHSs}
We call a symmetric real-valued map $k:\mathcal{X} \times \mathcal{X} \rightarrow \mathbb{R}$ a kernel, and assume $k$ is a positive-definite (PD) function according to the definition that follows. 
\medbreak

\begin{definition}
A continuous function $ k: \mathcal{X} \times \mathcal{X} \to \mathbb{R} $ is called positive-definite if for any set of pairwise-distinct sites $X = \{x_1,\dots,x_N\}$, with an arbitrary $N \in \mathbb{N}$, it holds that
\begin{equation}
    \sum_{i=1}^N \sum_{j=1}^N \alpha_i \alpha_j k(x_i,x_j) > 0 \nonumber
\end{equation}
for any set of weighting constants $\alpha_1,\dots,\alpha_N \in \mathbb{R} \backslash \{0\}$.
\end{definition}

Even though limiting our scope to positive-definite functions excludes certain kernels, this class encompasses powerful alternatives such as the squared-exponential, the inverse multiquadrics and the truncated power function \citep{wendland2004}. The first of the three is known to have the universal approximation property whenever $\mathcal{X}$ is compact \citep{micchelli2006universal}.
\medbreak

\begin{myRemark}
For clarity purposes, we denote by $K \in \mathbb{R}^{N\times N}$ the constant matrix that has kernel evaluations as elements, i.e., $k(x_i,x_j)$ at its $i$th row and $j$th column for $x_i,x_j \in X$. Moreover, $K_{Xx} : \mathcal{X} \rightarrow \mathbb{R}^N$ denotes the column vector \textit{function} $x \mapsto \begin{bmatrix}k(x_1,x) & \dots & k(x_N,x)\end{bmatrix}^\top$, and $K_{xX}$ simply represents its transpose.
\end{myRemark}

Given a kernel $k$, we denote the associated uniquely determined reproducing kernel Hilbert space (RKHS) by $\mathcal{H}$. The one-to-one correspondence between $k$ and $\mathcal{H}$ is guaranteed by the Moore–Aronszajn theorem \citep{aronszajn1950theory}. Each element $g\in\mathcal{H}$ is a map from $\mathcal{X}$ to $\mathbb{R}$ assuming the form of a weighted sum of kernels $g = \sum_{i \in \Omega_g} \alpha_i k(x_i, \cdot)$, where the index set $\Omega_g$ of $g$ can also be countably infinite, in which case the limit interpretation of the series applies. $\mathcal{H}$ is equipped with the inner product $\langle g,f \rangle _\mathcal{H} \, = \sum_{i\in \Omega_g} \sum_{j\in \Omega_f} \alpha_i \beta_j k(x_i,x_j)$ and the induced norm is $\Vert g \Vert_{\mathcal{H}} := \sqrt{\langle g,g \rangle_{\mathcal{H}}}$. Fixing any $x$ in $\mathcal{X}$, the corresponding evaluation functional $l_x: \mathcal{H} \rightarrow \mathbb{R}$ is bounded and takes any $g \in \mathcal{H}$ to its image, i.e., $l_x(g) = g(x) = \langle g, k(x,\cdot)\rangle_\mathcal{H}$, being linked to the reproducing property. Suppose that $g$ has a finite expansion in terms of $N_g$ kernel functions. Due to the reproducing and basic inner product properties, it holds that
\begin{align}
    \Vert g \Vert_\mathcal{H}^2 %&= \langle g, g \rangle_\mathcal{H} \\
    & = \left\langle \sum_{i=1}^{N_g}\alpha_i k(x_i, \cdot) , \sum_{i=1}^{N_g}\alpha_i k(x_i, \cdot) \right\rangle_\mathcal{H} \\
    & = \sum_{i=1}^{N_g} \sum_{j=1}^{N_g} \alpha_i \alpha_j k(x_i,x_j) \\
    & = \alpha^\top K \alpha \label{eq.normAlpha}
\end{align}
where $\alpha := \begin{bmatrix}\alpha_1 & \dots & \alpha_{N_g}\end{bmatrix}^\top$ gathers all weights.
\medbreak

\begin{assumption}
Given a kernel $k$, we assume that the ground-truth $f$ belongs to its RKHS, $\mathcal{H}$. Additionally, an upper bound for its norm $\Vert f \Vert_{\mathcal{H}} \le \Gamma$ is available.
\label{as.gtBound}
\end{assumption}

Establishing any form of guarantee is clearly impossible if no assumptions are made on the unknown map $f$. The availability of an upper bound $\Gamma$ is also assumed in the works \cite{srinivas2012information,koller2018learning,hashimoto2020learning}. It is beyond the scope of this paper to present various ways to construct bounds for $\rknorm{f}$. We nevertheless illustrate how this quantity can be estimated from perfect evaluations of $f$ in Section~\ref{sec:numericalExperiments}. The following function is defined as a last introductory step, which will later play an important role in our error estimates.
\medbreak

\begin{definition}
\label{def.powerFun}
The power function is the real-valued map $P_X(x) = \sqrt{k(x,x) - K_{xX}K^{-1}K_{Xx}}$.
\end{definition}

Throughout most of the document, $D$ is assumed to be fixed. For this reason, we drop the dependence that $P_X(x)$ has on the data sites to ease the notation, writing simply $P(x)$. Two main properties of this function will be exploited herein: 
\vspace{-0.2cm}
\begin{equation}
    P(x) \geq 0, \text{ for any } x \in \mathcal{X} \nonumber
\end{equation} 
\vspace{-0.5cm}
\begin{equation}
    P(x_n) = 0, \text{ for any } x_n \in X \nonumber
\end{equation}
\noindent \noindent which follow from rewriting the power function in a Lagrange form as shown in \cite[Sec.~11.1]{wendland2004}.

\section{Crafting models}
\label{sec:crafting}

We restrict our attention to models $s: \mathcal{X} \rightarrow \mathbb{R}$ built as a weighted sum of kernels that are centered at the data locations
\begin{equation}
    s(x) = \sum_{n=1}^N \alpha_n k(x_n,x)  = \alpha^\top K_{Xx}\, .
    \label{eq.basicmodel}
\end{equation}
Solutions to a number of optimal fitting problems have this form as discussed next. Since the number of functions $k$ and their centers have already been defined, constructing a model is equivalent to deciding the $\alpha$ coefficients.

\subsection{The noise-free case}

In the absence of noise ($\bar{\delta} = \mathbf{0}$), the labels in $D$ perfectly represent $f$. We can then solve the approximation problem by finding an $s \in \mathcal{H}$ such that the evaluations $s(x_i)$ match $f(x_i) =: f_{x_i}$ for all points in $D$. This can be posed as the variational problem
\begin{subequations}
\begin{align}
    \bar{s} = \argmin_{s \in \mathcal{H}} & \quad \rknorm{s}^2 \\
    \text{subj. to} & \quad s(x_n) = f_{x_n} \\
                & \quad \forall n = 1,\dots,N \nonumber
\end{align}
\label{eq.interpol}%
\end{subequations}
\noindent in which the objective favors low-complexity solutions, measured by the function space norm $\rknorm{\cdot}$. %This property will be central to derive prediction-error bounds for the resulting model $\bar s$.

Thanks to the optimal recovery property (see \cite[Thm~13.2]{wendland2004} or \cite[Thm.~ 3.5]{kanagawa2018gaussian}), it is known that out of all elements $s \in \mathcal{H}$ capable of interpolating the dataset, a minimizer for the above problem exists and assumes the form \eqref{eq.basicmodel}. The solution $\bar{s}$ can be therefore found by simply solving the linear system of equations $K \alpha = f_X$ for $\alpha$, where $f_X = \begin{bmatrix}f(x_1) & \dots & f(x_N)\end{bmatrix}^\top$. Given the PD property of the kernel $k$ and Assumption~\ref{as.pairwiseDistinct}, $K$ is positive-definite and hence invertible. Therefore, $\alpha = K^{-1}f_X$ and the unique optimizer of \eqref{eq.interpol} is
\begin{equation}
    \bar{s}(x) = f_X^\top K^{-1} K_{Xx}
    \label{eq.interpolModel}
\end{equation}
Because of \eqref{eq.normAlpha}, we see that its norm can be expressed in terms of the data values as $\rknorm{\bar{s}}^2 = f_X^\top K^{-1} f_X$.
\medbreak

\begin{myRemark}
\label{rem:normEst}
It holds that $\rknorm{\bar{s}} \leq \rknorm{f}$ independently of the number of samples in $D$. This stems from $f$ being the solution to \eqref{eq.interpol} when the equality constraints are imposed for all $x \in \mathcal{X}$.
\end{myRemark}

A first inequality can be obtained for the model $\bar{s}$ with the aid of the previous remark. This is a known but not very disseminated result, which tightens the more widely spread bound \cite[Thm.~11.4]{wendland2004}. The proof we give here is important to help build an intuition on how the RKHS norm measures the complexity of a function.
\medbreak

\begin{myProposition}
\label{prop.boundNoiseFree}
Assume that the dataset $D$ is not affected by noise, i.e., $\bar{\delta} = \mathbf{0}$ and $ y = f_X$. Under Assumptions~\ref{as.pairwiseDistinct} to \ref{as.gtBound}, the interpolating model $\bar{s}$ admits the error bound
\begin{equation}
    |\bar{s}(x) - f(x)| \leq P(x) \, \sqrt{\Gamma^2 - \rknorm{\bar{s}}^2}
    \label{eq.boundNoiseFree}
\end{equation}
for any $x \in \mathcal{X}$, where $f$ is the unknown ground-truth and $\rknorm{\bar{s}}^2 = f_X^\top K^{-1}f_X$.
\end{myProposition}

\begin{proof}
Let $x \in \mathcal{X}$ be a fixed query point, which is not in $D$. Denote by $\bar{s}_+$ the function of the form \eqref{eq.basicmodel} interpolating all known points $f_X$ in $D$ and the unknown value $f_x := f(x)$. We then have
\begin{subequations}
\begin{align}
    \rknorm{\bar{s}_+}^2 = &
    \begin{bmatrix} f_X \\ f_x \end{bmatrix}^\top 
    \begin{bmatrix} K \quad & K_{Xx} \\ K_{xX} & K_{xx} \end{bmatrix}^{-1}
    \begin{bmatrix} f_X \\ f_x \end{bmatrix} \nonumber \\[3pt]
    = & 
    \begin{bmatrix} f_X \\ f_x \end{bmatrix}^\top 
    \begin{bmatrix} K^{-1}  & \mathbf{0} \\ \mathbf{0} & 0 \end{bmatrix}
    \begin{bmatrix} f_X \\ f_x \end{bmatrix} + P^{-2}(x)
    \begin{bmatrix} f_X \\ f_x \end{bmatrix}^\top 
    \begin{bmatrix} K^{-1} K_{Xx} \\ -1 \end{bmatrix}
    \begin{bmatrix} K^{-1} K_{Xx} \\ -1 \end{bmatrix}^\top
    \begin{bmatrix} f_X \\ f_x \end{bmatrix} \nonumber \\[3pt]
    = & \, \rknorm{\bar{s}}^2 + P^{-2}(x) (\bar{s}(x) - f_x)^2  \nonumber \\[3pt]
   \leq & \, \Gamma^2 \nonumber 
\end{align}
\end{subequations}
where the second equality follows from the matrix inversion lemma, and the inequality follows from Remark~\ref{rem:normEst}. Finally, the last two lines imply $|\bar{s}(x) - f(x)| \leq P(x) \sqrt{\Gamma^2 - \rknorm{\bar{s}}^2}$. 

If, on the other hand, the query point $x$ belongs to the dataset $D$, the bound evaluates to zero and thus it holds tightly.
\end{proof}

One can also arrive at \eqref{eq.boundNoiseFree} by starting from the inequality $|\bar{s}(x) - f(x)| \leq P(x) \, \rknorm{f - \bar{s}}$ \cite[Eq.~9]{fasshauer2011positive} and noting that $\rknorm{f - \bar{s}}^2 = \rknorm{f}^2 - 2 \langle \bar s,f \rangle_{\mathcal{H}} + \rknorm{\bar s}^2 = \rknorm{f}^2 - 2 \rknorm{\bar s}^2 + \rknorm{\bar s}^2 \leq \Gamma^2 - \rknorm{\bar{s}}^2$. Nevertheless, the proof we provided gives more insight on how the norm of a model $\bar s$ might grow after the addition of a new data-point. More specifically, this only happens if the new value $f_x$ differs from what the model was previously predicting $\bar s(x)$.

Through Proposition~\ref{prop.boundNoiseFree}, evaluations of $f$ for every $x \in \mathcal{X}$ can be bounded according to $f^{\text{min}}(x) \le f(x) \le f^{\text{max}}(x)$ with $f^{\text{min}}(x) = \bar s(x) - P(x) \sqrt{\Gamma^2 - \rknorm{\bar s}^2}$ and $f^{\text{max}}(x) = \bar s(x) + P(x) \sqrt{\Gamma^2 - \rknorm{\bar s}^2}$. Proposition~\ref{prop:rec} below establishes that the interval containing the ground-truth is non-growing after the addition of a new data-point.
\medbreak

\begin{myProposition}
\label{prop:rec}
Let $x$ be any fixed query point in the domain $\mathcal{X}$. Let $Z$ be an augmented set of distinct data-sites, i.e., $Z = X\cup \{z\}$, $z\in \mathcal{X}, z\not\in X$. Then we have
\begin{equation}
    [f_Z^{\min}(x), f_Z^{\max}(x)] \subseteq [f_X^{\min}(x), f_X^{\max}(x)]
\end{equation}
\end{myProposition}
\begin{proof}
See Appendix \ref{pr:lemrec}.
\end{proof}

Intuitively, this favorable property states that augmenting the dataset $D$ with any new pair $(x,y) \in \mathcal{X}\times \mathbb{R}$ (while still satisfying Assumption~\ref{as.pairwiseDistinct}) either preserves or sharpens the inequality \eqref{eq.boundNoiseFree}. This holds everywhere in the domain.

\subsection{Kernel ridge regression analysis}
\label{subsec:krr}
To tackle the approximation problem in the presence of measurement noise, a compromise between fitting the data and rejecting uninformative fluctuations has to be found. One of the most standard tools used to achieve this balance is kernel ridge regression (KRR), in which the unconstrained problem
\begin{equation}
    s^\ast = \argmin_{s \in \mathcal{H}} \ \frac{1}{N}\sum_{n=1}^N \left( y_n-s(x_n) \right)^2 + \lambda \rknorm{s}^2
\label{eq.krr}
\end{equation}
is solved, yielding the KRR model $s^\ast$. In \eqref{eq.krr}, the regularization weight $\lambda \in \mathbb{R}_{\geq0}$ dictates the aforementioned balance: $\lambda=0$ produces a plain interpolant due to our assumptions on $k$ and $D$, whereas increasing values of $\lambda$ lead to a surrogate model that fits the dataset while avoiding abrupt variations. See \cite{scandella2020note} for a recent study on the numerical properties of the problem.

Thanks to the representer theorem, this infinite-dimensional functional problem over $\mathcal{H}$ can be recast as an equivalent finite-dimensional one (see the pivotal work \cite{scholkopf2001generalized}, and \cite{diwale2018generalized} for a recent generalization). A closed-form solution for \eqref{eq.krr} can then be obtained through this reformulation and is given by (see e.g. \cite[Thm.~3.4]{kanagawa2018gaussian})
\begin{equation}
    s^\ast(x) = \alpha^{\ast \top} K_{Xx}
    \label{eq.krrsol}
\end{equation}
with optimal weights $\alpha^\ast = (K+N\lambda I)^{-1}y$. Let $c$ denote the vector of values attained by $s^\ast$ at the data locations $X$, i.e., $c_n := s^\ast(x_n), n = 1,\dots,N$. The regressor then satisfies $K \alpha^\ast = c \Rightarrow \alpha^\ast = K^{-1}c$. From the latter and \eqref{eq.normAlpha}, the norm can be also expressed in the convenient quadratic form $\rknorm{s^\ast}^2 = c^\top K^{-1}c$, where $c = K(K+N\lambda I)^{-1} y$.

Before establishing the KRR prediction-error bound, we first need to analyze how noise can perturb the norm of an interpolant. To this end, consider the next result.
\medbreak

\begin{myLemma}
\label{lem.noisyNorm}
Let Assumptions~\ref{as.pairwiseDistinct} and \ref{as.noiseModel} hold. Let moreover $\bar{s}(x) = f_X^\top K^{-1} K_{Xx}$ be the model interpolating the noise-free values $f_X$, and $\tilde{s}(x) = y^\top K^{-1} K_{Xx}$ the model interpolating the noisy values $y$. Then
\begin{equation}
    \nabla \leq \rknorm{\tilde{s}}^2 - \rknorm{\bar{s}}^2 \leq \Delta
\end{equation}
where $\Delta$ denotes the maximum and $\nabla$ the minimum of $(- \delta^\top K^{-1} \, \delta + 2 \, y^\top K^{-1} \, \delta)$ subject to $|\delta| \leq \bar{\delta}$.
\end{myLemma}

\begin{proof} 
It follows from expanding $\rknorm{\tilde s}^2$ as $\rknorm{\bar s}^2$ plus a perturbation term, and recalling the definitions of $\Delta$ and $\nabla$.
\end{proof}

Whereas calculating $\Delta$ amounts to solving a convex optimization problem since it is the maximum of a strictly concave function, evaluating $\nabla$ is not as straightforward. Still, the quantity $\nabla$ is not employed in our expressions. Bounding the error associated with the KRR predictions is then possible through the following inequality.
\medbreak

\begin{myTheorem}
\label{thm.krrBound}
Let $N$ be the number of data-points, $\bar{\delta} \in \mathbb{R}^N_{>0}$ the noise bound, and $\lambda$ the regularization constant. Under Assumptions~\ref{as.pairwiseDistinct} to \ref{as.gtBound}, the KRR model $s^\ast$ admits the error bound
\begin{align}
|s^\ast(x)-f(x)| & \leq \, P(x) \, \sqrt{ \Gamma^2 + \Delta - \rknorm{\tilde{s}}^2} + \bar{\delta}^\top \, \vert K^{-1}K_{Xx} \vert + \left| \, y^\top \left(K + \frac{1}{N \lambda} \, KK \right)^{-1} K_{Xx}\right|
\label{eq.boundKRR}
\end{align}
for any $x \in \mathcal{X}$, where $f$ is the unknown ground-truth, $\Delta = \max_{|\delta| \leq \bar{\delta}} (- \delta^\top K^{-1} \delta + 2 y ^\top K^{-1} \delta)$, and $\rknorm{\tilde{s}}^2 = y^\top K^{-1} y$.
\end{myTheorem}

\begin{proof}
Recall that $\bar{s}(x) = f_X^\top K^{-1} K_{Xx}$ and $y = f_X + \delta$. Predictions given by $s^\ast(x)$ can be decomposed as
\begin{subequations}
\begin{align}
    s^\ast(x) 
    = \; & y^\top(K+N \lambda I)^{-1}K_{Xx} \\[5pt]
    = \; & f_X^\top(K+N \lambda I)^{-1}K_{Xx} + \delta^\top(K+N \lambda I)^{-1}K_{Xx} \\[5pt]
    = \; & f_X^\top K^{-1}K_{Xx} - f_X^\top (K+\frac{1}{N \lambda}KK)^{-1}K_{Xx} + \delta^\top(K+N\lambda I)^{-1}K_{Xx} \label{eq.woodburry1} \\
    = \; & f_X^\top K^{-1}K_{Xx} - y^\top (K+\frac{1}{N\lambda}KK)^{-1}K_{Xx} + \delta^\top \left[ (K+\frac{1}{N\lambda}KK)^{-1}  + (K+N\lambda I)^{-1} \right] K_{Xx} \\
    = \; & \bar{s}(x) - y^\top (K+\frac{1}{N\lambda}KK)^{-1}K_{Xx} + \delta^\top K^{-1}K_{Xx} \label{eq.woodburry2}%
\end{align}
\end{subequations}
where \eqref{eq.woodburry1} and \eqref{eq.woodburry2} both follow from Woodbury's matrix identity. For compactness, let $Q = (K + \frac{1}{N\lambda}KK)$. The error norm can therefore be upper bounded by
\begin{subequations}
\begin{flalign}
    |s^\ast(x) - f(x)| &= |\bar{s}(x) - f(x) + \delta^\top K^{-1}K_{Xx} - y^\top Q^{-1}K_{Xx}| \\[4pt]
    & \leq |\bar{s}(x) - f(x)| + |\delta^\top K^{-1}K_{Xx} - y^\top Q^{-1}K_{Xx}|  \label{eq.dueToTriangle} \\[4pt]
    & \leq |\bar{s}(x) - f(x)| +  \bar{\delta}^\top \, \vert K^{-1}K_{Xx} \vert + |y^\top Q^{-1}K_{Xx}| \label{eq.tightThing} \\[4pt]
    & \leq P(x) \sqrt{\Gamma^2 - \rknorm{\bar{s}}^2} +  \bar{\delta}^\top \, \vert K^{-1}K_{Xx} \vert + |y^\top Q^{-1}K_{Xx}| \label{eq.almostLast}%
\end{flalign}
\end{subequations}
in which the triangle inequality was used to obtain \eqref{eq.dueToTriangle}. Note that \eqref{eq.tightThing} \textit{tightly} bounds \eqref{eq.dueToTriangle}, i.e., $\exists \delta: |\delta| \leq \bar\delta$ such that both expressions are equal. Finally, \eqref{eq.almostLast} is due to Proposition~\ref{prop.boundNoiseFree}. Thanks to Lemma~\ref{lem.noisyNorm}, we know that $\rknorm{\tilde{s}}^2 - \Delta \leq \rknorm{\bar{s}}$, concluding the proof.
\end{proof}

Let us have a closer look at the bound \eqref{eq.boundKRR}. Firstly, it is consistent with the interpolating noise-free case, i.e., if $\bar{\delta} = \mathbf{0}$ and $\lambda \rightarrow 0$, \eqref{eq.boundKRR} converges to \eqref{eq.boundNoiseFree}. Secondly, the constant $\Delta$ was introduced since we do not have access to $\rknorm{\bar{s}}^2$. Calculating it requires solving a box-constrained quadratic program (QP) for it is the maximization of a concave objective. The maximum is moreover \textit{independent} of the query point $x$. % and, as mentioned before, can be approximated from above by $\bar{\Delta} = 2 \, \bar{\delta}^\top \vert K^{-1} y \, \vert$ if needed.
As a compelling alternative, note that $P(x) \, \Gamma$ can be used as a replacement for $P(x) \, \sqrt{ \Gamma^2 + \Delta - \rknorm{\tilde{s}}^2}$ in \eqref{eq.boundKRR}, being a simple consequence of \eqref{eq.boundNoiseFree}. By doing that, we observed experimentally that little conservativeness is introduced, while avoiding the need to compute $\Delta$.

Approaching the problem from another perspective, we analyze now a technique that enjoys a convenient low-norm solution, which in turn leads to simplified error bounds.

\subsection{$\varepsilon$-Support vector regression analysis}
%In this subsection, we formulate a new approximation problem and derive deterministic error bounds for its solution
We now aim at finding a minimum norm solution to the approximation problem under noise, while also fully exploiting the boundedness of $\delta$. This can be readily formulated as
\begin{subequations}
\begin{align}
    s^\star = \argmin_{s \in \mathcal{H}} & \quad \rknorm{s}^2 \\
    \text{subj. to} & \quad | s(x_n) - y_n| \leq \bar\delta_n \\
                & \quad \forall n = 1,\dots,N \nonumber
\end{align}
\label{eq:svr}%
\end{subequations}
where $\bar\delta_n$ is the $n$th element of the $\bar\delta$ vector. The mathematical program above can be interpreted as a $\varepsilon$-support vector regression (SVR) problem with hard margins \citep{schlkopf2018learning}.

With the help of an indicator function, the constraints above can be incorporated into the objective, permitting the use of the representer theorem once more \citep{scholkopf2001generalized}. The optimizer can be found by solving the simple quadratic program
\begin{subequations}
\begin{align}
    \alpha^\star = \argmin_{\alpha \in \mathbb{R}^N} & \quad \alpha^\top K \alpha \\
    \text{subj. to} & \quad | K \alpha - y| \leq \bar\delta 
\end{align}
\label{eq:svrweights}%
\end{subequations}
and setting $s^\star(x) = \alpha^{\star\top} K_{Xx}$,
where the attained values at the data sites are denoted by $d_n := s^\star(x_n)$, $n=1,\dots,N$, with $d=\begin{bmatrix} d_1 & \dots & d_N\end{bmatrix}^{\top}=K\alpha^{\star}$. 
Since the ground-truth is a feasible solution to \eqref{eq:svrweights}, the order $\Vert s^{\star}\Vert_{\mathcal{H}} \le \Vert f \Vert_{\mathcal{H}} \le \Gamma$ holds. Furthermore, we highlight that
\begin{equation}
\label{eq:svrlebar}
   \rknorm{s^{\star}}\le \rknorm{\bar s} 
\end{equation}
which also follows from the fact that the function $\bar s$, interpolating the noise free samples, is contained in the search space of \eqref{eq:svrweights}. This enables us to bound the prediction-error associated with the SVR model $s^{\star}$.
\medbreak

\begin{myTheorem}
\label{thm.svrBound}
Let $\bar{\delta} \in \mathbb{R}^N_{>0}$ be the noise bound. Under Assumptions~\ref{as.pairwiseDistinct} to \ref{as.gtBound}, the SVR model $s^\star$ admits the error bound 
\begin{align}
    \vert s^{\star}(x) - f(x) \vert & \le \; P(x)\sqrt{ \Gamma^2- \rknorm{s^{\star}}^2} + \label{eq.SVRbound}\, 
    \bar\delta^\top \vert K^{-1}K_{Xx} \vert + \vert (d-y)^{\top} K^{-1}K_{Xx} \vert 
\end{align}
for all $x\in \mathcal{X}$, where $f$ is the unknown ground-truth and $\rknorm{s^{\star}}^2 = d^\top K^{-1} d$.
\end{myTheorem}

\begin{proof}
Observe that $d-f_X = d-y +\delta$, with $\vert \delta \vert \le \bar\delta$ by Assumption~\ref{as.noiseModel} and where $d-y$ is known. We get
\begin{align}
    \vert s^{\star}(x) - f(x) \vert &= \vert d^\top K^{-1}K_{Xx} - f(x) \vert \label{eq:svrbound1} \\[4pt]
    & \le \vert f_X^\top K^{-1}K_{Xx} - f(x) \vert \label{eq:svrbound2} + \vert  (d-f_X)^\top K^{-1}K_{Xx}\vert \\[4pt]
    & \le P(x) \sqrt{\Gamma^2 - \rknorm{\bar s}^2} + \vert (d-y + \delta)^\top K^{-1}K_{Xx} \vert \label{eq:svrbound3} \\[4pt]
    & \le P(x) \sqrt{\Gamma^2 - \rknorm{s^\star}^2} + \vert (d-y)^\top K^{-1}K_{Xx} \vert +\bar\delta^\top \vert K^{-1}K_{Xx}\vert \label{eq:svrbound5}
\end{align}
Where \eqref{eq:svrbound2} follows from the triangle inequality, \eqref{eq:svrbound3} from Proposition \ref{prop.boundNoiseFree} and the observation, and \eqref{eq:svrbound5} from \eqref{eq:svrlebar}, the triangle inequality and the noise bound.
\end{proof}

We notice again that this bound is consistent with the noise-free case: for $\bar \delta = \mathbf{0}$, it holds that $d_n = f(x_n)$ and $ \rknorm{s^{\star}} = \rknorm{\bar s}$, so we recover the bound in Proposition~\ref{prop.boundNoiseFree}.

\section{Discussion}
\label{sec.discussion}

\subsection{Comparing KRR and SVR}

Whether a KRR or SVR model should be used in a given context depends on a number of practical considerations. Strengths and weaknesses of each approach are examined next.

\textit{Model computation:}
The KRR model can be directly constructed from a dataset $D$ as in \eqref{eq.krrsol}, since \eqref{eq.krr} admits a closed-form solution. The SVR problem \eqref{eq:svr}, on the other hand, does not have an explicit solution in terms of the data, but it requires the user to solve \eqref{eq:svrweights}.

\textit{Hyperparameters:}
The choice of the regularization parameter $\lambda$ in KRR is not straightforward, and should be guided by the knowledge one has on $\delta$. A badly chosen regularizer impacts not only the model quality, but also the size of the error bounds \eqref{eq.boundKRR}. In the Gaussian processes field, hyperparameter learning (including the kernel constants) is typically done through maximal likelihood estimation \citep{williams2006gaussian}; in the non-Bayesian kernel literature, different forms of cross validation are usually employed \citep{duan2003evaluation}. In contrast with KRR, no hyperparameter is involved in the SVR alternative, making it more suitable to scenarios where nothing is known about the possible noise realization.

\textit{Bound computation:}
To compute $\Delta$ in \eqref{eq.boundKRR}, an optimization problem has to be solved. Since $\Delta$ does not depend on the query point location $x$, this has to be done only once for a fixed dataset $D$ and noise bound $\bar \delta$. As opposed to it, bounding the prediction error of an SVR model can be done directly as the inequality \eqref{eq.SVRbound} only depends on given, fixed quantities.

\textit{Error bounds magnitude:}
Differences in magnitude and shape between the error bounds \eqref{eq.boundKRR} and \eqref{eq.SVRbound} are mainly due to their last absolute value terms. They are related to the difference of the respective kernel model and the interpolant $\tilde s$. In principle, choosing $\lambda$ small in the KRR setting would lead to tighter bounds. Nevertheless, this would also deteriorate the nominal model $s^*$ performance as it would fit noise rather than filtering it. An appropriate $\lambda$ is the key to attain a smooth nominal predictor while keeping the bounds small. The SVR is constrained to stay close to $\tilde s$ at the sample locations by definition, maximizing smoothness within the available margins. Although in our experiments they performed similarly and none of the bounds strictly encompassed the other, the KRR one seems to be slightly tighter on average for a well chosen regularization weight $\lambda$.

\subsection{The effects of incorporating new data}
\label{subsec:newdata}
The problem of improving the reconstruction quality of a surrogate model is central in approximation theory \citep{iske2000optimal,de2003optimal,de2005near}. Three main tools are usually employed to analyze it: the Lebesgue function $ \mathcal{L}(x) := \Vert K^{-1}K_{Xx}\Vert_1$, the fill distance and the separation distance, respectively defined as
\begin{align*}
    h_{D,\mathcal{X}} & := \sup_{x \in \mathcal{X}} \min_{x_n \in D} \Vert x - x_n \Vert_2  \\
    q_{D} & := \min_{\substack{x_i, x_j \in D \\ x_i \neq x_j}} \, \frac{1}{2} \, \Vert x_i - x_j \Vert_2 
\end{align*}
Notice that, if the noise bound is uniform across all samples, then $\bar\delta^\top \vert K^{-1}K_{Xx} \vert$ present both in \eqref{eq.boundKRR} and \eqref{eq.SVRbound} simplifies to $\mathcal{L}(x)$ times the bounding constant. Due to this term, it is not guaranteed that the bounds will shrink everywhere
after the addition of new data-points at arbitrary locations. It is known that a new datum is most benign when it minimizes $h_{D,\mathcal{X}}$ while \textit{not reducing} $q_{D}$. Balancing these two constants is an issue commonly referred to as the \textit{uncertainty principle} \cite{fasshauer2011positive}. A key and simple advice is to use a uniformly or quasi-uniformly distributed dataset, which not only favors the bound shrinkage, but also controls the increase of the kernel matrix $K$ condition number \cite{hangelbroek2010kernel,diederichs2019improved}. This suggests that if the data at hand are highly scattered, a pre-processing stage is highly recommended, possibly dropping points that are too close to each other as they could lead to numerical instabilities.

Despite the statements made above, and using the same arguments as in \cite[Sec.~4.1]{hashimoto2020learning}, one sees that the space to where the unknown ground-truth is confined is non-increasing with the addition of any new datum. In more contrete terms, denote the right-hand side of \eqref{eq.boundKRR} by $e(x)$, and let $W_D(x) := \{y \in \mathbb{R} \, | \, s^\ast(x) - e(x) \leq y \leq s^\ast(x) + e(x) \}$ be the interval function that bounds the value of $f(x)$ for all $x \in \mathcal{X}$. Let $D^+$ be the dataset augmented with one (pairwise-distinct) point. Whereas it is not guaranteed that $W_{D^+}(x) \subset W_D(x)$, clearly $f$ must satisfy $f(x) \in W_D(x) \cap W_{D^+}(x)$, so that the confinement space is always non-increasing, i.e., total information gathered about the unknown function increased. The same observations are clearly true for the SVR model as well.

\subsection{On the connections between KRR and Gaussian processes}
\label{subsec:GPcomparison}

Consider without loss of generality a Gaussian process setting with a null prior mean function. Let $\sigma^2_{\delta}$ denote the variance of the Gaussian measurement noise. Then, the GP conditional expectation defines \textit{exactly the same map} from $\mathcal{X}\rightarrow \mathbb{R}$ as the KRR solution $s^\ast$ with a regularization parameter $\lambda = \sigma^2_\delta / N$ \cite[Sec.~3]{kanagawa2018gaussian}. Indeed, GP posterior means have precisely the same form as \eqref{eq.krrsol}. The disparity between both methodologies is in the way their hypotheses spaces are defined. On one hand, $f$ is assumed to follow a distribution governed by the covariance kernel $k$, and on the other, $f$ is a static member of the RKHS of the kernel $k$. In other words, in the former case $f$ is a stochastic process realization in a suitable probability space, and in the latter, $f$ is a fixed map. Bounds derived for each model are therefore very different in nature: GP results draw probabilistic limits for their sample paths, whereas KRR results bound any function in $\mathcal{H}$, including the ground-truth \cite{kanagawa2018gaussian}.

In adopting the perspective presented in this paper, one might be concerned with the restrictiveness of working only in the space $\mathcal{H}$ \citep{lederer2019uniform}. It has been shown that, whereas a GP mean belongs to its reproducing kernel Hilbert space, sample paths fall outside of it almost surely as they are `more complex' maps \cite[Sec.~4]{kanagawa2018gaussian}. Nonetheless, given any continuous function $g$, the set $\mathcal{H}$ associated with for instance the squared-exponential kernel has at least one member that is arbitrarily close to $g$|that is, $\mathcal{H}$ is dense in the class of continuous functions \citep{micchelli2006universal}. Models $s$ in $\mathcal{H}$ enjoy therefore the so called universal approximation property, which renders their representation capabilities equal to many families of widely used neural networks.

\section{Numerical experiments}
\label{sec:numericalExperiments}

Two examples are presented here to illustrate the behavior of the established inequalities in different conditions\footnote{The code to reproduce the examples is available at \texttt{https://github.com/emilioMaddalena/DetErrBnd}.}.
First we compare the KRR and SVR approaches to each other, and to the deterministic error bound recently proposed in \cite{hashimoto2020learning}\footnote{The results were derived from the well-known paper \cite{srinivas2012information}.}. In the cited work, the authors proposed a GP-like surrogate model and established deterministic error bounds also exploiting an RKHS norm estimate $\Gamma$. The expression is reproduced here using our notation:
\begin{equation}
    |s^\diamond(x)-f(x)| \leq \sigma(x) \, \sqrt{\Gamma^2 - y^\top (K + \tilde \delta^2 I)^{-1}y + N} \label{eq:japan}
\end{equation}
where $s^\diamond$ is their model, $\sigma^2(x) = k(x,x)-K_{xX} (K + \tilde \delta^2 I)^{-1} K_{Xx}$, and $\tilde \delta$ is a \textit{necessarily uniform} bound on the noise. A second example is also discussed, focusing exclusively on the KRR case. The negative effects of having highly scattered data are highlighted and a simple way to handle them is shown.

\begin{figure}[t]
    \centering
    \includegraphics[width=0.49\linewidth]{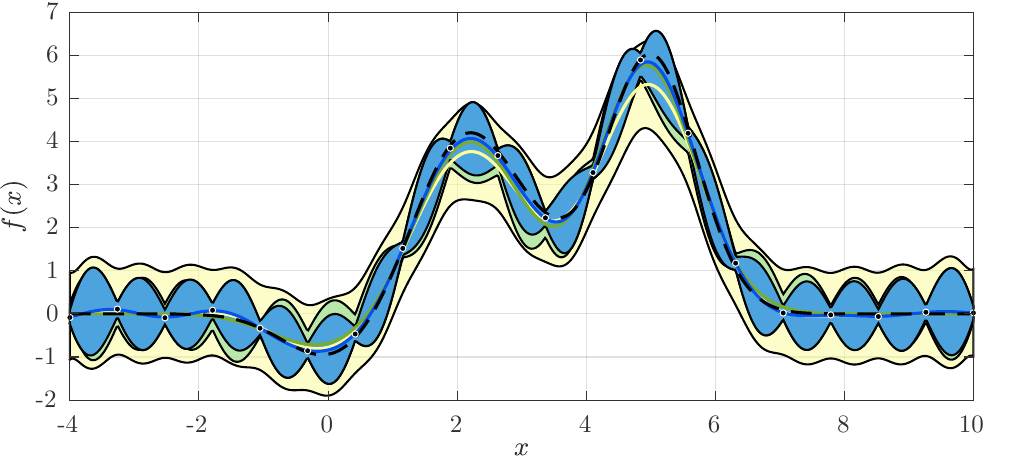} %\\[3pt]
    \includegraphics[width=0.49\linewidth]{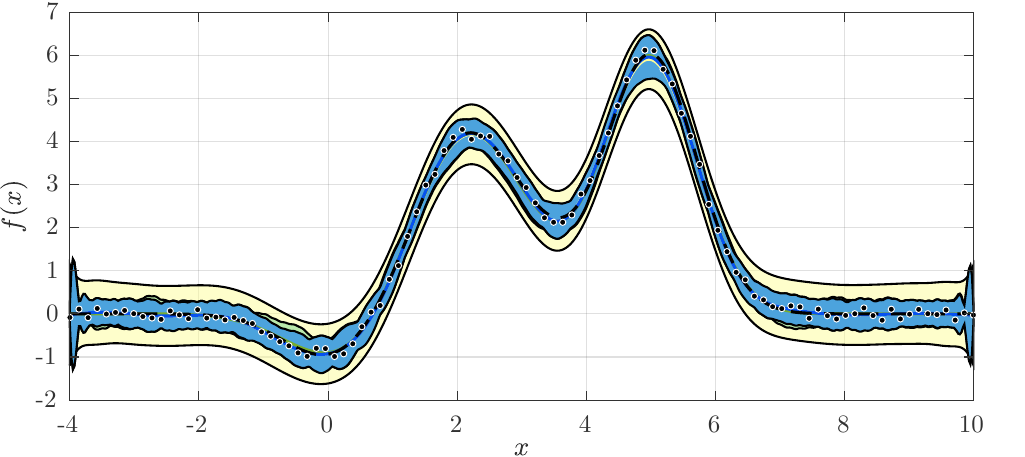}
    \caption{Ground-truth (\textbf{- -}), KRR (\textcolor{coolBlue}{\textbf{|}}), SVR (\textcolor{coolGreen}{\textbf{|}}), and the alternative proposed in \cite{hashimoto2020learning} (\textcolor{coolYellow}{\textbf{|}}). The error bounds are depicted using the same colors of their respective models, and were computed for $N=20$ (top) and $N=100$ (bottom) samples. The noisy data-points are shown as black circles.}
    \label{fig:example1}
\end{figure}

\begin{figure*}[t!]
    \centering
    \vspace{0.15cm}
    \hspace{0.02\linewidth} \includegraphics[width=0.95\linewidth]{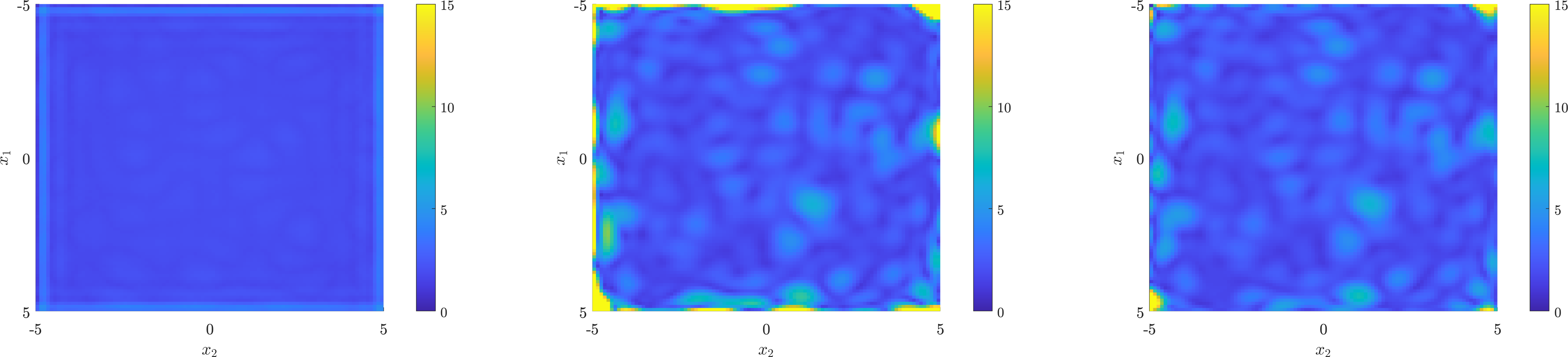}\\[5pt]
    \includegraphics[width=0.98\linewidth]{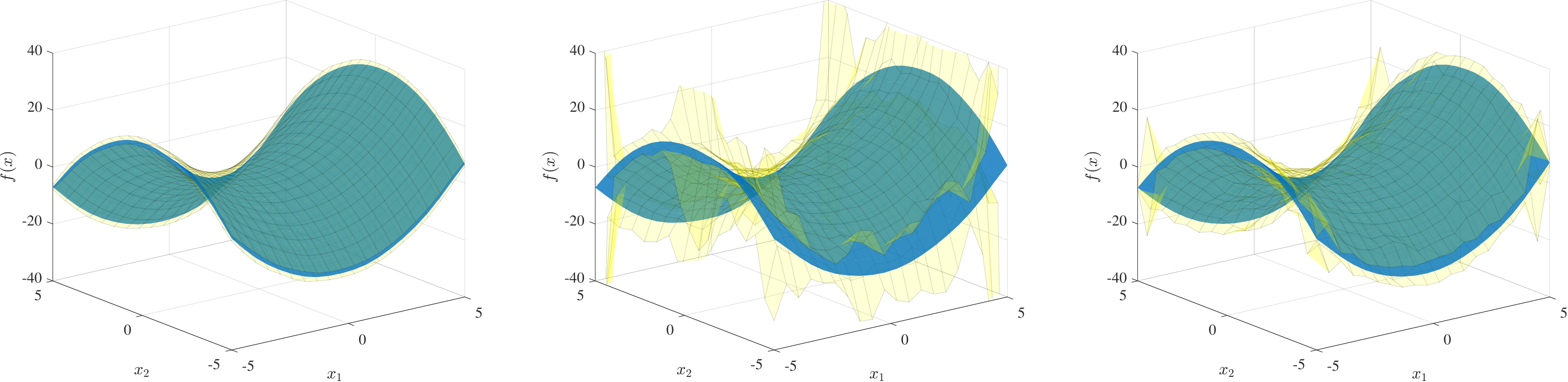}
    \caption{Error bound magnitude $e(x)$ (colormap, top) and KRR regressors along with their prediction envelopes (bottom) for three scenarios: (left) $N=625$ points distributed in a grid, (center) $N=625$ sampled from a uniform random distribution, and (right) $N=625$ sampled from a uniform random distribution plus $44$ points collected from the edges.}
    \label{fig:3Dcomp}
\end{figure*}

\subsection{A comparison among three alternatives}

Let $k$ be the isotropic squared-exponential kernel function 
\begin{equation}
    k(x,x_n) = \exp \left(-\frac{\Vert x-x_n\Vert^2_2}{2\ell^2}\right) 
\end{equation}
with lengthscale $\ell = 0.707$, and consider a ground-truth function $f:\mathcal{X} \to \mathbb{R}$ with a known kernel expansion $f(x) = - k(x,0) + 3.5 \, k(x,2) + 1.6 \, k(x,3) + 6 \, k(x,5)$, which leads us to $\rknorm{f} = 7.49$. Surrogate models were built based on an overestimate $\Gamma = 9$, and two datasets sampled from $\mathcal{X} = \{ x \in \mathbb{R} \, | \, -4 \leq x \leq 10 \}$, affected by a uniformly bounded noise $\vert \delta_n \vert \leq 0.15$, for all $n$. The ridge regression (with $\lambda = 0.001$), support vector regression and the model in \cite{hashimoto2020learning} were computed and are shown in Figure~\ref{fig:example1}. In terms of nominal predictions, the three approaches yielded similar results. The SVR model however was able to filter the existing noise best (for this specific KRR choice of $\lambda$). As for the bounds, the technique proposed in \cite{hashimoto2020learning} encompassed the KRR and SVR areas almost everywhere in both scenarios, therefore being more conservative. In this particular example, this was due to two reasons. Firstly, recall Definition~\ref{def.powerFun}, and notice that the power function $P(x)$ will always evaluate to a number smaller than $\sigma(x)$ in \eqref{eq:japan} due to $(K+\tilde \delta^2 I) \succ K$. Secondly, \eqref{eq:japan} has a direct dependence on the number of samples $N$, which is not present in the other two bounds. The uncertainty regions defined by the KRR and SVR models were similar to a large extent in shape and magnitude, indicating that both are equally suitable modeling tools. When the number of samples was increased from $N=20$ to $N=100$, all bounds tended to shrink: the average thickness of the prediction envelope was reduced from $2.14$ to $1.41$ (yellow), $1.20$ to $0.73$ (blue), and $1.35$ to $0.74$ (green). However, one can see a growth of the blue and green bounds at the extremes of the domain, when $N$ was increased. This is a common effect and was mainly caused by the Lebesgue-related term present in \eqref{eq.boundKRR} and \eqref{eq.SVRbound} (see Subsection~\ref{subsec:newdata}).
% thickness: GP 2.14, KRR 1.2, SVR 1.35
% thickness: GP 1.41, KRR 0.73, SVR 0.75

\subsection{The KRR bounds in 2 dimensions}

Consider the dynamics of the Tinkerbell chaotic system first coordinate $f(x) = x_1^2 - x_2^2 + 0.8 \, x_1 - 0.6 \, x_2$ on the domain $\mathcal{X} = \{ x \in \mathbb{R}^2 \, | \, \begin{bmatrix}-5 & -5\end{bmatrix}^\top \leq x \leq \begin{bmatrix}5 & 5\end{bmatrix}^\top \}$. With a slight abuse of notation, $x_1$ and $x_2$ denote the first and second components of $x$. Two training datasets of $N = 625$ points were collected: one forming a perfect grid across the domain, and one drawn randomly from a uniform distribution. Bounded measurement noise $|\delta_n| \leq 0.5$ for all $n$ was considered in both cases. The squared-exponential kernel was chosen and the lengthscale $\ell = 1.62$ was determined by maximizing the resulting log-likelihood for a sensible variance value. $\Gamma$ was estimated by collecting noiseless evaluations $f_X$ of the ground-truth and determining the norm of the associated interpolant; a final value of $\Gamma = 196.1$ was adopted after the use of a safety factor (see Remark~\ref{rem:normEst}). The KRR model \eqref{eq.krrsol} with $\lambda = 1\times 10^{-5}$ was used to reconstruct $f$. The final surrogate functions $s^\ast(x)$ and their error bounds $e(x)$ (the right-hand side of inequality \eqref{eq.boundKRR}) are shown in Figure~\ref{fig:3Dcomp}. 

As can be seen from the first two surfaces, the nominal KRR models were very similar despite being trained with differently distributed datasets. The bounds were tight and uniform under the evenly spaced samples, whereas they behaved badly under the scattered ones, showing high peaks especially at the border of the domain. We stress that the scale was held constant (from $0$ to $15$) across $e(x)$ plots for visualization purposes, but the attained values were higher in the completely saturated yellow regions as indicated by the prediction envelope below it. Notice nevertheless that the center part of the error bounds remained relatively tight. Finally, the random dataset was augmented with $44$ points collected from the domain boundary, and the results are presented in the rightmost plots. Incorporating these extra points was enough to significantly dampen the bounds increase not only at the borders, but also in internal regions. We observed an average error bound of $2.10$ for the grid sampling, an increase to $3.60$ after randomizing the sample locations, and a reduction to $2.66$ in the final case, providing thus a clear overall improvement.

\section{Final remarks}
\label{sec:conclusion}

Deterministic prediction-error bounds were provided for two classes of popular non-parametric kernel techniques: kernel ridge regression and $\varepsilon$-support vector regression. In our setting, we considered bounded measurement noise and assumed that the ground-truth function belongs to the RKHS induced by a known positive-definite kernel. We believe that our uncertainty bounds can be employed in a number of different scenarios such as in the deterministic certification of machine learning algorithms. The inequalities established herein are centered around each of the models; a pertinent question is whether the maps $f \in \mathcal{H}$, $\rknorm{f} \leq \Gamma$ alone also admit (non-trivial) bounds for their point-wise evaluations. As a last observation, the matter of hyperparameter selection is still open: what specific metrics could be considered when learning them in our bounded noise setting? 

\appendix

\section{Proof of Proposition~\ref{prop:rec}}
\label{pr:lemrec}
In the following, we only prove $f^{\max}_X(x)\ge f^{\max}_Z(x)$, the inequality for $f^{\min}$ follows from the same arguments, which proves the interval containment.
To get the interval at a point $x\in \mathcal{X}$, we consider the sets $X, \bar X=X\cup\{x\}, Z=X\cup\{z\}$ and $W=Z\cup\{x\}$. Additionally, we denote the interpolant of $f_X$ by $ s_X$ and follow this convention for the other sets. We observe the following norm identities, derived as in the proof of Proposition \ref{prop.boundNoiseFree}
\begin{align}
    \rknorm{s_W}^2 &= \rknorm{s_{\bar X}}^2 + P_{\bar X}^{-2}(z)(s_{\bar X}(z)-f_z)^2 \\[4pt]
    &= \rknorm{s_{X}}^2 + P_X^{-2}(x)(s_{X}(x)-f_x)^2 + P_{\bar X}^{-2}(z)(s_{\bar X}(z)-f_z)^2 \\[4pt]
    &= \rknorm{s_{Z}}^2 + P_Z^{-2}(x)(s_{Z}(x)-f_x)^2 \\[4pt]
    &\le \Gamma^2
\end{align}
This allows us to write $f^{\max}_Z(x)$ in two different ways
\begin{align}
    f^{\max}_Z(x) &=s_{Z}(x) + P_Z(x)\sqrt{\Gamma^2-\rknorm{s_{Z}}^2} \\[4pt]
    &= s_{X}(x) + P_X(x)\sqrt{\Gamma^2-\rknorm{s_{X}}^2 - P_{\bar X}^{2}(s_{\bar X}(z)-f_z)^2} \label{eq:way2}
\end{align}
From \eqref{eq:way2}, we observe
\begin{align}
f^{\max}_Z(x) \le s_{X}(x) + P_X(x)\sqrt{\Gamma^2-\rknorm{s_{X}}^2} = f^{\max}_X(x)
\end{align}
using the positivity of the power function.
$\QED$

% Acknowledgments---Will not appear in anonymized version
\acks{This work received support from the Swiss National Science Foundation under the Risk Aware Data-Driven Demand Response project (grant number 200021 175627) and CSEM's Data Program.}

\bibliography{refs}

\end{document}